\documentstyle[12pt,amsmath,amssymb]{article}
\begin{document}

\tolerance=5000

\def\be{\begin{equation}}
\def\ee{\end{equation}}
\def\bea{\begin{eqnarray}}
\def\eea{\end{eqnarray}}
\def\nn{\nonumber \\}
\def\e{{\rm e}}

\begin{titlepage}

\begin{center}
\Large
{\bf Where new gravitational physics comes from: M-theory?}

\vfill

\normalsize

\large{ 
Shin'ichi Nojiri$^\spadesuit$\footnote{Electronic mail: nojiri@nda.ac.jp, 
snojiri@yukawa.kyoto-u.ac.jp} and 
Sergei D. Odintsov$^{\heartsuit\clubsuit}$\footnote{Electronic mail:
 odintsov@ieec.fcr.es Also at TSPU, Tomsk, Russia}}

\normalsize

\vfill

{\em $\spadesuit$ Department of Applied Physics, 
National Defence Academy, \\
Hashirimizu Yokosuka 239-8686, JAPAN}

\ 

{\em $\heartsuit$ Institut d'Estudis Espacials de Catalunya (IEEC), \\
Edifici Nexus, Gran Capit\`a 2-4, 08034 Barcelona, SPAIN}

\ 

{\em $\clubsuit$ Instituci\`o Catalana de Recerca i Estudis 
Avan\c{c}ats (ICREA), Barcelona, SPAIN}

\end{center}

\vfill

\baselineskip=24pt
\begin{abstract}

It is suggested \cite{CDTT} that current cosmic acceleration arises 
due to modification of General Relativity by the terms with negative
powers of curvature. We argue that time-dependent
(hyperbolic) compactifications of $4+n$-dimensional gravity ( 
string/M-theory) lead to the effective 4d 
gravity which naturally contains such terms. The same may be achieved in 
the braneworld by the proper choice of the boundary action. Hence, such a
model which seems to eliminate the need for dark energy may have the
origin in M-theory.

\end{abstract}


\end{titlepage}

\noindent
{\bf 1. Introduction.} 
Recent astrophysical data from Ia supernovae\cite{R} 
and cosmic microwave radiation \cite{B} clearly indicate that 
current Universe is undergoing a phase of the accelerated expansion.
Despite many theoretical efforts it remains unclear what is the origin 
of such accelerated behaviour. Phenomenologically, the easiest possibility 
is to supply the standard FRW cosmology with cosmic fluid with large,
negative pressure (dark energy).
This dark energy is expected to carry most of the universe energy density.
Various candidates for dark energy were proposed: the cosmological
constant (for a review, see \cite{C}), scalar fields, extra dimensions,
etc. It is clear that such a list may be significally extended.
Unfortunatelly, neither of existing dark energy models is completely satisfactory.

In such a situation one can assume that cosmic speed-up is due to 
new gravitational physics which becomes important at very low
curvatures. In other 
words, the Einstein gravity should be modified 
providing the effective dark energy contribution at late times.
The model of this sort has been recently proposed in ref.\cite{CDTT}
(for discussion of related models, see also \cite{CCS}).
It is a combination of the term with the 
scalar curvature $R$ and the inverse of $R$. Note that other terms with
negative powers of curvature may be introduced as well. Such models may
have problems with late time production of small/zero curvature black
holes. The starting action is \cite{CDTT}\footnote{ More generally,
the theory which contains any negative power of $R$ is called the
modified gravity. There are astrophysical indications \cite{CCS} that
power of $1/R$ term should be fractional and close to 1.}:
\be
\label{R1}
S={1 \over \kappa^2}\int d^4 x \sqrt{-g}\left( R - {\mu^4 \over R}\right)\ .
\ee
By introducing an auxiliary field $A$, one may rewrite the above action 
(\ref{R1}):
\be
\label{R2}
S={1 \over \kappa^2}\int d^4 x \sqrt{-g}\left( R - 2\mu^2 A + A^2 R\right)\ .
\ee
Using the equation of motion for $A$,
we find the two actions (\ref{R1}) and (\ref{R2}) are equivalent. With the
redefinition of the auxiliary field $A$ by
$\e^{2\phi} = 1 + A^2$, the action (\ref{R2}) becomes:
\be
\label{R5}
S={1 \over \kappa^2}\int d^4 x \sqrt{-g}\left( \e^{2\phi}R \mp 2\mu^2 
\sqrt{1 - \e^{2\phi}}\right)\ .
\ee
Rescaling the metric tensor $g_{\mu\nu}$ by 
$g_{\mu\nu}\to\e^{-2\phi}g_{\mu\nu}$,
 the action (\ref{R5}) is transformed as
\bea
\label{R7}
S&=&{1 \over \kappa^2}\int d^4 x \sqrt{-g}\left( R 
 - 6 g^{\mu\nu}\partial_\mu \phi \partial_\nu \phi \right.\nn
&&\left. \mp 2\mu^2 \e^{-4\phi} \sqrt{1 - \e^{2\phi}}\right)\ .
\eea
Then the action  (\ref{R1}) is equivalent to the Einstein 
action coupled with the scalar field $\phi$ with potential 
$2\mu^2 \e^{-4\phi} \sqrt{1 - \e^{2\phi}}$.
As it has been demonstrated in ref.\cite{CDTT} (see also \cite{CCS})
the late-time cosmic acceleration can really arise due to such small
modification of the General Relativity (GR) by $1/R$ or by $1/R^m$ terms.

Of course, giving rise the effective dark energy,
such a modification of General Relativity looks also completely ad hoc.
The deep problem remains: where these curvature corrections come from?
In order to find some way towards to solution of above problem it may be 
right time to compare the successful cosmic acceleration models where new
gravitational physics appears. In particulary, recently the accelerating
cosmologies from higher dimensional gravity with hyperbolic,
time-dependent compactification were found \cite{TW,ohtan} (for related
discussion and extensions, see \cite{EG,others,CHNOW}). Note that such
solutions 
were also obtained earlier in a different context, as $S$-branes in
refs.\cite{sbrane}. The influence of hyperbolic extra dimensions to 4d
cosmology has been studied earlier in refs.\cite{trodden}.  
It is remarkably, that actually string/M-theory
compactification gives rise to such accelerating cosmologies \cite{TW} 
because the standard no-go theorems are not applied for such
compactifications (for example, where these theorems are not applied to
time-dependent torus compactification, see \cite{costa}).
In the present Letter, we argue that time-dependent, hyperbolic
compactifications of string/M-theory \cite{TW,ohtan,others,CHNOW} or
braneworld scenario actually
may reproduce the modifications of GR of the sort 
proposed in ref.\cite{CDTT}. This indicates (but, unfortunately, does
not prove) that string/M-theory may be
the source of new gravitational physics!


\noindent
{\bf 2. Time-dependent (hyperbolic) compactification.}
Let us consider now if the model (\ref{R1}) can be obtained from (time-dependent) 
compactification. We start from $4+n$-dimensional spacetime, whose metric 
is given by
\be
\label{R8}
ds^2 = \sum_{\mu,\nu=0,1,2,3}g_{\mu\nu} dx^\mu dx^\nu + \e^{2\phi(x^\mu)}
\sum_{i,j=1}^n \tilde g_{ij} d\xi^i d\xi^j\ .
\ee
For simplicity, one may assume the metric $\tilde g_{ij}$ expresses the 
Einstein manifold, where the Ricci tensor $\tilde R_{ij}$ constructed from 
$\tilde g_{ij}$ is proportional to $\tilde g_{ij}$: $\tilde R_{ij}=k \tilde g_{ij}$.
Here $k$ is a constant. When $n=1$, $k$ always vanishes ($k=0$). When $n\geq 3$, the 
 above metric is given as the solution of 
$n$-dimensional
Euclidean Einstein equation. When $n=2$, since 2d Einstein equation is
trivial, 
in the conformal gauge
the above condition for Ricci tensor is the Liouville equation.

Under the above assumptions, the $4+n$ dimensional Einstein action can be
rewritten as \footnote{For simplicity, instead of $p$-form contribution which
 is typical in M-theory we consider cosmological constant contribution.
Then, the above action does not literally correspond to
superstring/M-theory. However, it reflects the main features of such
unified models in many respects.}
 \bea
\label{R10}
S_{4+n}&=&{1 \over \kappa^2}\int d^{4+n} x \sqrt{-g^{(4+n)}}\left(R^{(4+n)} - \Lambda\right) \nn
&=& {V_n \over \kappa^2}\int d^4 x \sqrt{-g} \e^{n\phi}\left( R \right. \nn
&& \left. - \Lambda + n(n-1)g^{\mu\nu}
\partial_\mu \phi \partial_\nu \phi + nk\e^{-2\phi}\right)\ .
\eea
Here $V_n$ is the volume of the $n$-dimensional manifold whose metric tensor is 
given by $\tilde g_{ij}$. 
If we rescale the 4-dimensional metric $g_{\mu\nu}$ by
$g_{\mu\nu}\to \e^{-n\phi}g_{\mu\nu}$ ,
the action (\ref{R10}) can be rewritten as
\bea
\label{R12}
S_{4+n}&=& {V_n \over \kappa^2}\int d^4 x \sqrt{-g}\left( R 
 - {n(n+2) \over 2}g^{\mu\nu}\partial_\mu \phi \partial_\nu \phi \right.\nn 
&& \left. - \e^{-n\phi}\Lambda+ nk\e^{-(n+2)\phi}\right)\ .
\eea
This is Einstein gravity coupled with the scalar field $\phi$, whose 
potential is $\e^{-n\phi}\Lambda - nk\e^{-(n+2)\phi}$. 
The obtained action is almost identical with that in \cite{EG} where 
the starting action contains $p$-form field. Instead of the $p$-form field, 
 the cosmological constant is included in the original action (\ref{R10}). 
The qualitative structure of the action (\ref{R12}) is the same as
that in \cite{EG}. 
Since the potential is given by
\be
\label{R13a}
V(\phi)= \e^{-n\phi}\Lambda - nk\e^{-(n+2)\phi}\ ,
\ee
if $\Lambda>0$ and $k<0$ (this case is more attractive due to stability),
the potential is monotonically decreasing
function 
with respect to $\phi$ and vanishes at $\phi\to + \infty$. Then as in \cite{EG}, 
for large $\phi$ with very large kinetic energy with negative 
velocity $\dot\phi<0$, the universe starts decelerated expansion. At some
point 
$\phi=\phi_0$, the velocity vanishes $\dot \phi=0$, and after that 
 an accelerated expansion begins. 

We now further rescale the 4-dimensional metric $g_{\mu\nu}$ in the action (\ref{R12}) by
$g_{\mu\nu}\to \e^{\phi\sqrt{n(n+2) \over 3}}g_{\mu\nu}$,
then the action  (\ref{R12}) can be rewritten as
\bea
\label{R15}
S_{4+n}&=& {V_n \over \kappa^2}\int d^4 x \sqrt{-g}\left( \e^{\alpha\phi}R \right. \nn
&&\left. - \e^{\phi\left(2\alpha -n\right)}\Lambda
+ nk\e^{\phi\left(2\alpha - n -2\right)}\right)\ .
\eea
Here $\alpha\equiv \sqrt{n(n+2) \over 3}$.
Since the kinetic term for $\phi$ vanishes, one may regard $\phi$ as an
auxiliary field. 
By the variation of the action (\ref{R15}) with respect to $\phi$, we obtain
\bea
\label{R17}
R=f(\phi)&\equiv& \left(2 - {n \over \alpha}\right)\e^{\left(\alpha -n\right)\phi}\Lambda \nn
&& - \left(2 - {n + 2 \over \alpha}\right)nk \e^{\left(\alpha - n -2\right)\phi}\ .
\eea
Solving (\ref{R17}) with respect to $\phi$ when $n\neq 0$, one may delete
$\phi$ 
in (\ref{R15}). The resulting action looks like
\be
\label{R18}
S_{4+n}= {V_n \over \kappa^2}\int d^4 x \sqrt{-g} F(R)\ .
\ee
With no matter and for the Ricci tensor $R_{\mu\nu}$ being
covariantly constant, the equation of motion is:
\be
\label{R19}
0=2F(R) - RF'(R)\ ,
\ee
which is the algebraic equation with respect to $R$. Solving 
(\ref{R19}) with 
respect to $R$, one may find $\phi$ from (\ref{R17}).  

When $n=6$, Eq.(\ref{R17}) can be exactly solved as
\be
\label{RE1}
\e^\phi=\sqrt{2R \over \Lambda}\ ,
\ee
and we obtain
\be
\label{RE2}
F(R)=-{\Lambda \over 2R} + 6k\ .
\ee
The strange property of above formula is that Einstein term does not
appear there. There may be several mechanisms behind such a behaviour:

\noindent
1. Our $4+n$-dimensional gravity is not effective string theory even
for $n=6$. One can expect that in full superstring theory where there 
are contributions from other fields the Einstein term will be restored.

\noindent
2. Another proposal could be related with the compactification under the
discussion. For instance, the physical compactification could be somehow
different, say, like $H_n/\Gamma$.

When $|k|\ll |\Lambda|$, Eq.(\ref{R17}) can be solved as 
\be
\label{R20}
\e^\phi \sim \left\{{\alpha \over 2\alpha - n}{R \over \Lambda}\right\}^{1 \over \alpha -n}\ .
\ee
and one gets
\be
\label{R21}
F(R)= f_0^{(n)} \left({R \over \Lambda}\right)^{2 + {n \over \alpha - n}}
+ f_1^{(n)} \left({k \over \Lambda}\right)
\left({R \over \Lambda}\right)^{2 + {n - 2 \over \alpha - n}}
+ f_2^{(n)} \left({k \over \Lambda}\right)^2
\left({R \over \Lambda}\right)^{2 + {n - 4 \over \alpha - n}}
+ \cdots\ .
\ee
Here $f_i^{(n)}$'s are coefficients depending on $n$. Especially when $n=7$, 
which may correspond to M-theory, we numerically obtain 
\be
\label{R21b}
F(R)= f_0^{(7)} \left({R \over \Lambda}\right)^{-0.86}
+ f_1^{(7)} \left({k \over \Lambda}\right)
\left({R \over \Lambda}\right)^{-0.04}
+ f_2^{(7)} \left({k \over \Lambda}\right)^2
\left({R \over \Lambda}\right)^{0.77} + \cdots\ .
\ee
We again see that the terms with fractional negative powers of $R$ appear.
Such the action may produce the required cosmic acceleration. However, the
Einstein term is induced only approximately. (Note 
the astrophysical data suggest that power of $R$ in Einstein gravity 
may vary around 1 with 10\% accuracy!) As in previous case of $n=6$ one
can expect that in full M-theory with account of $p$-form contribution 
the correct Einstein term may be induced within existing cosmological
bounds. Or, once more the physical compactification which corresponds to
our evolving FRW universe is somehow different from the simple case under
consideration. 
 
On the other hand, when $|k|\gg |\Lambda|$, Eq.(\ref{R17}) can be solved as 
\be
\label{R22}
\e^\phi \sim \left\{-{\alpha \over 2\alpha - n -2 }{R \over nk}\right\}^{1 \over \alpha -n-2}\ .
\ee
and 
\be
\label{R23}
F(R)\propto R^{2 + {n+2 \over \alpha - n-2}}\ .
\ee
We should note $2 + {n \over \alpha - n}=\infty$ and $2 + {n+2 \over \alpha - n-2}>0$ 
when $n=1$ and $2 + {n \over \alpha - n}>0$ for $0<n<1$ and 
$2 + {n \over \alpha - n}<0$ for $n>1$. 
We should also note $2 + {n+2 \over \alpha - n-2}=0$ for $n=6$, 
$2 + {n+2 \over \alpha - n-2}>0$ for $n>6$ and 
$2 + {n+2 \over \alpha - n-2}<0$ for $n<6$. When $n\geq 2$, there effectively appears the
negative power 
of the curvature in 4d action. 

Several more remarks about Eq.(\ref{R17}) are in order. 
We should note that $2 - {n \over \alpha}$ is always positive. On the 
other hand, $2 - {n+2 \over \alpha}<0$ when $n<6$, $2 - {n+2 \over \alpha}>0$ when $n>6$, 
and $2 - {n+2 \over \alpha}=0$ when $n=6$. 
$n=\alpha<n+2$ when $n=1$ and $\alpha < n<n+2$ when $n\geq 2$. 

We first consider $n=1$ case, where $k=0$ and $\alpha=1$. Then
(\ref{R17}) gives 
$R=\Lambda$. One cannot solve (\ref{R17}) with respect to $\phi$. 
Consider $n\geq 2$ case.  
If $k>0$ and $n<6$ ($k<0$ and $n>6$), $f(\phi)\to +\infty$ 
($f(\phi)\to -\infty$) when $\phi\to -\infty$ and $f(\phi)\to 0$ when $\phi\to +\infty$. 
If the scalar curvature is large, $F(R)$ in the action behaves as (\ref{R23}) 
and if the scalar curvature is small, $F(R)$ behaves as (\ref{R21}). 
When $n=6$, it is described by (\ref{R21}),that is $F(R)\propto {1 \over R}$. 
When $n\geq 6$, for both,
large or small curvature, the action is described by the positive power of $R$. 
On the other hand, when $n<6$, the action corresponds to the positive power of $R$ when 
$R$ is small but to the negative power of $R$ when $R$ is large. This is
the inverse of the behaviour in (\ref{R1}). 

When $\Lambda>0$, $k<0$, and $n<6$ ($\Lambda<0$, $k>0$, and $n<6$), 
or  $\Lambda>0$, $k>0$, and $n>6$ ($\Lambda<0$, $k<0$, and $n>6$)
$f(\phi)$  (\ref{R17}) is a monotonically decreasing (increasing) function. 
When $\Lambda,k>0$ and $n<6$ ($\Lambda,k<0$ and $n<6$), 
or $\Lambda>0$, $k<0$ and $n>6$ ($\Lambda<0$, $k>0$ and $n>6$), 
$f(\phi)$ has a zero at finite $\phi$: 
$f(\phi_0)=0$. If $R$ is small, there is also a solution $\phi\sim \phi_0$
in 
(\ref{R17}). Then the action (\ref{R15}) takes the form of the Einstein
action with 
cosmological constant.

Thus, we showed that there are similarities between effective
4d gravitational action  (after time-dependent (hyperbolic)
compactification
of M-theory) and the starting $1/R$ action.


\noindent
{\bf 3. Product hyperbolic compactifications.} One may consider the product 
compactification more general than (\ref{R8}) 
as in \cite{CHNOW}.:
\bea
\label{R26}
ds^2 &=& \sum_{\mu,\nu=0,1,2,3}g_{\mu\nu} dx^\mu dx^\nu + \e^{\beta_1\phi(x^\mu)}
 \sum_{i,j=1}^n g^{(1)}_{ij} d\xi^i d\xi^j \nn
&& + \e^{\beta_2\phi(x^\mu)} \sum_{I,J=1}^m g^{(2)}_{IJ} d\xi^i d\xi^j\ .
\eea
(As a further generalization,  
 some function $\sigma(\phi)$ may be introduced in the exponent. Then, 
 the negative power of the scalar curvature could appear more naturally.)
Starting from $4+n+m$-dimensional Einstein action and
 rescaling the 4-dimensional metric $g_{\mu\nu}$ by
$g_{\mu\nu}\to \e^{-{n\beta_1 + m\beta_2 \over 2}\phi}g_{\mu\nu}$, 
the classical action becomes
\bea
\label{R29}
&& S_{4+n+m}= {V_n V_m \over \kappa^2}\int d^4 x \sqrt{-g} \left\{ R 
 - \Lambda \e^{-{n\beta_1 + m\beta_2 \over 2}\phi}\right. \nn
&& - \left( {\beta_1^2 n + \beta_2^2 m \over 4} 
+ {\left(\beta_1 n + \beta_2 m\right)^2 \over 8}\right)g^{\mu\nu}
\partial_\mu \phi \partial_\nu \phi \nn
&& \left. + nk^{(1)}\e^{-{(n+2)\beta_1 + m\beta_2 \over 2}\phi} 
+ mk^{(2)}\e^{-{n\beta_1 + (m+2)\beta_2 \over 2}\phi}\right\}.
\eea
We now further rescale the 4-dimensional metric $g_{\mu\nu}$ in the action (\ref{R29}) by
$g_{\mu\nu}\to \e^{\phi\sqrt{{\beta_1^2 n + \beta_2^2 m \over 6} 
+ {\left(\beta_1 n + \beta_2 m\right)^2 \over 12}}}g_{\mu\nu}$.
The action (\ref{R29}) can be rewritten as
\bea
\label{R31}
&& S_{4+n+m} = {V_n V_m \over \kappa^2}\int d^4 x \sqrt{-g}\left( \e^{\gamma\phi}R \right. \nn
&& - \e^{\phi\left(2\gamma - {n\beta_1 + m\beta_2 \over 2}\right)}\Lambda   
+ nk^{(1)}\e^{\phi\left(2\gamma - {(n+2)\beta_1 + m\beta_2 \over 2}\right)} \nn
&& \left. + nk^{(2)}\e^{\phi\left(2\gamma - {n\beta_1 + (m+2)\beta_2 \over 2}\right)}\right)\ .
\eea
Here
\be
\label{R32}
\gamma\equiv \sqrt{{\beta_1^2 n + \beta_2^2 m \over 6} 
+ {\left(\beta_1 n + \beta_2 m\right)^2 \over 12}}\ .
\ee
Let us consider the situation that
\bea
\label{R33}
3\gamma&=&(n+2)\beta_1 + m \beta_2\ ,\nn
\tilde\gamma &\equiv&  -2\gamma + {n\beta_1 + (m+2)\beta_2 \over 2} \gg \gamma\ .
\eea
It is chosen $\Lambda=0$. Then the action (\ref{R31}) has the following form
\bea
\label{R34}
S_{4+n+m}&=& {V_n V_m \over \kappa^2}\int d^4 x \sqrt{-g}\left( \e^{\gamma\phi}R \right.\nn
&& \left. + nk^{(1)}\e^{{\gamma \over 2}\phi}+ nk^{(2)}\e^{-\tilde\gamma\phi}\right)\ .
\eea
By the variation over $\phi$, one finds
\be
\label{R35}
R=\tilde f(\phi)\equiv -{n\gamma k^{(1)} \over 2}\e^{-{\gamma \over 2}\phi} 
 - n\tilde\gamma k^{(2)}\e^{\left(-\tilde\gamma-\gamma\right)\phi}\ .
\ee
Then if $k^{(1)},k^{(2)}>0$ ($k^{(1)},k^{(2)}<0$), $\tilde f(\phi)$ is a monotonically 
increasing (decreasing) function. When $\phi\to +\infty$, the first term in 
$\tilde f(\phi)$ becomes dominant and $\tilde f(\phi) \to 0$. When $R$ is
small $\e^{{\gamma \over 2}\phi}\sim {1 \over R}$, and 
\be
\label{R37}
S_{4+n+m}\sim \int d^4x \sqrt{-g}\left({1 \over R}\right)\ .
\ee
On the other hand, when $\phi\to -\infty$, the second term in $\tilde f(\phi)$ becomes 
dominant and goes to infinity ($\left|\tilde f(\phi)\right|\to
\infty$). If $R$ is large, one gets
$\e^{{\gamma \over 2}\phi}\sim R^{-{\gamma \over \tilde\gamma + \gamma}}$, 
and 
\be
\label{R39}
S_{4+n+m} \sim \int d^4x \sqrt{-g}R^{1-{\gamma \over \tilde\gamma + \gamma}}\ .
\ee
Then in the limit $\tilde\gamma \gg \gamma$, Eqs.(\ref{R37}) and (\ref{R39}) reproduce 
the behaviour in (\ref{R1}). 
How realistic is a condition (\ref{R33})?
The first equation $3\gamma=(n+2)\beta_1 + m \beta_2$  (\ref{R33}) 
 can be always solved with respect to ${\beta_1 \over \beta_2}$. 
It is difficult, however, to realize the second condition 
$\tilde\gamma \gg \gamma$  (\ref{R33}) since 
$\tilde\gamma - {\gamma \over 2}=\beta_1$. It is more easier to get the
fractional negative power of $R$.

In order to avoid the above difficulty, we now consider further generalization of 
the product compactification like 
\be
\label{R42}
ds^2 = \sum_{\mu,\nu=0,1,2,3}g_{\mu\nu} dx^\mu dx^\nu + \sum_{l=1}^L\e^{\beta_l\phi(x^\mu)}
 \sum_{i,j=1}^{n_l} g^{(l)}_{ij} d\xi^i d\xi^j \ .
\ee
Then starting from $D=4+\sum_{l=1}^L n_l$-dimensional Einstein gravity, 
instead of (\ref{R31}), one introduces the parameter
\be
\label{R44}
\hat\gamma\equiv \sqrt{{\sum_{l=1}^L n_l\beta_l^2 \over 6} 
+ {\left(\sum_{l=1}^L n_l\beta_l\right)^2 \over 12}}\ .
\ee
The Ricci tensor $R^{(l)}_{ij}$ constructed from $g^{(1)}_{ij}$ is
 assumed to be  
proportional to $g^{(1)}_{ij}$: $R^{(1)}_{ij}=k^{(l)}g^{(1)}_{ij}$.
 With number of higher-dimensional parameters, by adjusting them, we may
satisfy the condition 
corresponding to (\ref{R33}):
\be
\label{R45}
3\hat\gamma= \sum_{l=1}^L n_l\beta_l + \beta_1 \ ,\quad
\tilde\gamma \equiv  -2\hat\gamma + {\sum_{l=1}^L n_l\beta_l + 2\beta_2  \over 2} 
\gg \hat\gamma\ .
\ee
Finally, putting $\Lambda=k^{(3)}=k^{(4)}=\cdots = k^{(L)}=0$ the
effective  action 
(\ref{R34}) behaves as (\ref{R37}) for small $R$ and as (\ref{R39}) 
for large $R$.   


\noindent
{\bf 4. Discussion.} It is shown that time-dependent (hyperbolic)
 compactification of $4+n$-dimensional gravity leads to the effective 4d
gravity action which 
naturally contains the (fractional) negative powers of curvature.
However, there are problems to get such an effective 4d gravity which also
contains the Einstein term on the same time. In our explicit example 
of ten- and eleven-dimensional gravity, the corresponding power of $R$ is 0
or about 0.8 (which is still below the admitted value between 0.9-1.1).
There may be several explanations of such behaviour.

\noindent
1. The full string/M-theory calculation with the account of all fields 
may presumably improve the situation.

\noindent
2. The question which compactification is physical one remains so far
unresolved in full string/M-theory.

\noindent
3. The ortodox point of view that results of second and third sections of
present work have nothing to do with string/M-theory may be also admitted.

\noindent
Nevertheless, the fact that negative power of $R$ effectively appears is
quite convincing.

It is also interesting that term like ${1 \over R}$  (\ref{R1}) might 
appear in the braneworld scenario \cite{RS} which is believed to be
related 
with string/M-theory too. 
Let the 3-brane is embedded into the 5d bulk space as in \cite{SMS}.
Let $g_{\mu\nu}$ be the metric tensor of the bulk space and $n_\mu$ be the unit vector 
normal to the 3-brane. Then the metric $q_{\mu\nu}$ induced on the brane
is $q_{\mu\nu}=g_{\mu\nu} - n_\mu n_\nu$.
Our starting action is: 
\be
\label{S00}
S=\int d^5 x \sqrt{-g}\left\{ {1 \over \kappa_5^2} R^{(5)} - 2\Lambda + \cdots \right\}
+ S_{\rm brane}(q)\ .
\ee
Neglecting the terms with the higher powers of the curvature, the 
effective Einstein equation is given by (for instance, \cite{SMS})
\bea
\label{S2d}
&& {1 \over \kappa_5^2} \left( R^{(4)}_{\mu\nu} - {1 \over 2} q_{\mu\nu}R^{(4)}\right) \nn
&=& - {1 \over 2}\left( \Lambda + {\kappa_5^2 \lambda^2 \over 6} \right) q_{\mu\nu} 
+ {\kappa_5^2 \lambda \over 6}\tau_{\mu\nu}\ .
\eea
Here $\lambda$ is the tension of the brane.
Then the corresponding effective 4d action follows
\be
\label{RR2}
S={1 \over \kappa_4^2}\int d^4 x\sqrt{-q}\left\{R^{(4)} - \Lambda_4\right\}
+ S_{\rm brane}(q)\ .
\ee
Here
\be
\label{RR3}
{1 \over \kappa_4^2}\equiv {6 \over \kappa_5^4 \lambda}\ ,\quad
\Lambda_4 \equiv \kappa_5^2
\left( \Lambda + {\kappa_5^2 \lambda^2 \over 6} \right) \ .
\ee
We now choose $S_{\rm brane}(q)$ to be an action of a dilaton gravity:
\be
\label{RR4}
S_{\rm brane}(q)= - {1 \over \kappa_4^2}\int d^4 x\sqrt{-q}\e^{2\phi}
\left\{R^{(4)} + 4 q^{\mu\nu}\partial_\mu \phi \partial_\nu \phi\right\}\ .
\ee
Under the conformal transformation $g_{\mu\nu}\to \e^\sigma g_{\mu\nu}$, 
with the choice $\sigma = \ln \left(1 + \e^\phi\right)$, 
the action (\ref{RR2}) with (\ref{RR4}) can be transformed as
\be
\label{RR8}
S = {1 \over \kappa_4^2}\int d^4 x\sqrt{-q}
\left\{ A^2\left(2-A\right)R^{(4)} + A\Lambda_4 \right\}\ .
\ee
Here $A\equiv 1+\e^{\phi}=\e^\sigma$. 
In (\ref{RR8}), the kinetic term of $\phi$ vanishes. $A$ can be regarded
as an auxiliary field. 
By the variation of the action with respect to $A$, we obtain
\be
\label{RR10}
\left(4A-3A^2\right)R^{(4)} + \Lambda_4 = 0\ ,
\ee
When $R$ is large, $A$ behaves as $A\to {4 \over 3}$ or ${\Lambda \over 2R}$.
For first branch $A\to {4 \over 3}$ the Einstein gravity is reproduced:
\be
\label{RR13}
S\to {1 \over \kappa_4^2}\int d^4 x\sqrt{-q}
\left\{ {32 \over 27}R^{(4)} + {4 \over 3}\Lambda_4 \right\}\ .
\ee
On the other hand, for second branch $A\to {\Lambda \over 2R}$, ${1 \over R}$-gravity 
is reproduced:
\be
\label{RR14}
S\to {1 \over \kappa_4^2}\int d^4 x\sqrt{-q}
\left\{{\Lambda^2 \over R}\right\}\ .
\ee
Therefore when $R$ is large, there are two branches: where gravity 
is the Einstein one or where ${1 \over R}$-gravity occurs. 
We may consider the limit that $R$ is small, where $A\to \pm \sqrt{\Lambda \over 3R}$. 
The 4d action (\ref{RR8}) behaves as ${1 \over \sqrt{R}}$-gravity:
\be
\label{RR16}
S\to {1 \over \kappa_4^2}\left(1 \pm {1 \over 3}\right)\Lambda 
\int d^4 x\sqrt{-q}\sqrt{\Lambda \over R}\ .
\ee
In the same way, changing the brane action (boundary terms in AdS/CFT) one
can get other negative
powers of the curvature.

Thus, there is some ground to beleive that gravitational
alternative to dark energy (which may be called the effective dark energy)
may be produced by the modification of GR which is dictated 
by string/M-theory. In such case, the mysterious cosmic fluid with 
negative pressure does not occur in the accord with the proposal
\cite{CDTT}. It is remarkable that the terms with positive higher
derivative curvature\footnote{The number of instabilities are predicted 
for original $1/R$ theory. Moreover, this theory seems non-realistic 
as equivalent scalar- tensor theory may not pass the simplest solar system 
tests. However, the addition of higher derivative terms like $R^2$ in
modified gravity 
may help to resolve the above instabilities and to make the scalar mass very heavy by simply tuning of the coefficient of $R^2$ term \cite{R2}.} 
like those in trace anomaly-driven inflation
\cite{star} (for a recent discussion in relation with dark energy, see
\cite{no}) may produce the early time inflation. These terms may also have
string/M-theory origin. Hence, GR is modified in a different way 
at early and late times. It remains to be the standard GR in intermediate
epoch only. The appearing new gravitational physics helps to realize both
phases: early time inflation and current cosmic acceleration. We presented
the arguments
that such modifications of GR may be predicted by M-theory.
However, the complete proof of this statement in full string/M-theory is
still missing.

\ 

\noindent
{\bf Acknowledgments.} The research is supported in part by the Ministry of
Education, Science, Sports and Culture of Japan under the grant n.13135208
(S.N.), DGI/SGPI (Spain) project BFM2000-0810 (S.D.O.), RFBR grant 03-01-00105
(S.D.O.) and LRSS grant 1252.2003.2 (S.D.O.).

\end{document}